\documentclass[10pt, conference]{IEEEtran}
\usepackage{waspaa25}

\ifCLASSINFOpdf
\else
   \usepackage[dvips]{graphicx}
\fi
\usepackage{url}
\usepackage{overpic}

\usepackage{graphicx}
\usepackage{amsmath}
\usepackage{hyperref}
\usepackage{amssymb}
\usepackage{float}
\usepackage{colortbl}
\usepackage{algorithm,algpseudocode}
\usepackage[font=small,skip=5pt]{caption}

\usepackage{bm} 

\usepackage{cleveref}

\title{Rethinking Non-Negative Matrix Factorization with Implicit Neural Representations}

\name{Krishna Subramani$^{1}$, Paris Smaragdis$^{1}$, Takuya Higuchi$^2$, Mehrez Souden$^2$}
\address{$^{1}$University of Illinois at Urbana-Champaign, $^{2}$Apple
}

\begin{document}

\maketitle

\begin{abstract}
Non-negative Matrix Factorization (NMF) is a powerful technique for analyzing regularly-sampled data, i.e., data that can be stored in a matrix. For audio, this has led to numerous applications using time-frequency (TF) representations like the Short-Time Fourier Transform. However extending these applications to irregularly-spaced TF representations, like the Constant-Q transform, wavelets, or sinusoidal analysis models, has not been possible since these representations cannot be directly stored in matrix form. In this paper, we formulate NMF in terms of learnable functions (instead of vectors) and show that NMF can be extended to a wider variety of signal classes that need not be regularly sampled.\footnote{Code: \url{https://github.com/SubramaniKrishna/in-nmf}}
\end{abstract}

\section{Introduction}
\label{sec:intro}

\begin{figure}
    \centering
    \includegraphics[scale = 0.45]{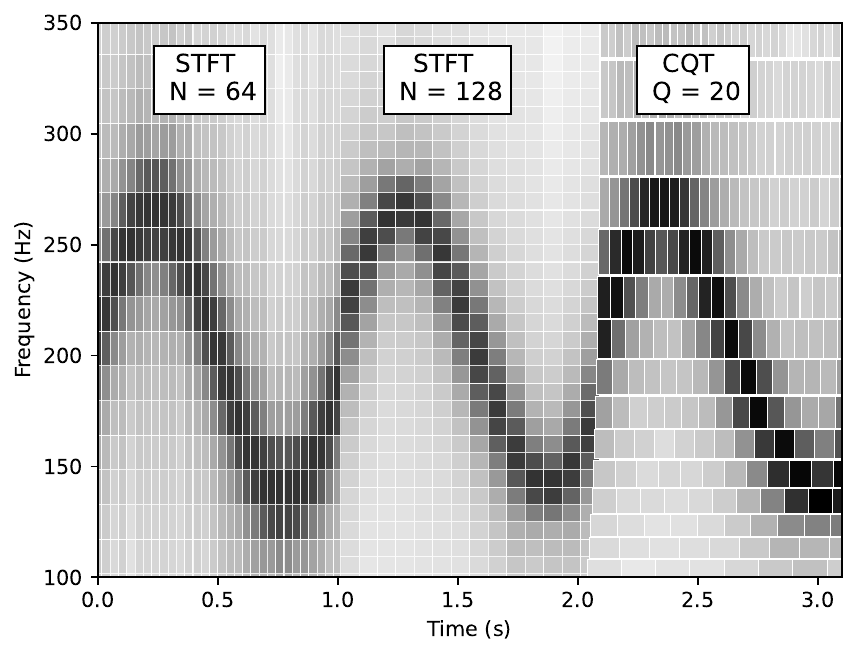}
    \caption{Changing the time and frequency resolution as we measure a signal. The initial small window size has a fine time resolution, while switching to a larger window afterwards exhibits better frequency resolution. At the end, we start using a multi-resolution decomposition where the low frequencies have better frequency resolution, while the high frequencies have better time resolution. The irregular tiling of the overall transformation above does not allow us to directly store this data as a matrix, which makes factorization operations on this data impossible.
    }
    \label{fig:stft_cqt}
\end{figure}

Consider a non-negative time-frequency representation for a frequency-modulated signal as shown in \autoref{fig:stft_cqt}. Unlike a conventional Short-Time Fourier Transform (STFT), this representation cannot be directly stored in a matrix because it is not regularly sampled across its two dimensions. As a result, this is not a representation directly compatible with any matrix factorization techniques. This paper will demonstrate an approach that allows us to factorize such irregular representations.

NMF approximates a non-negative matrix $\mathbf{V} \, \in  \mathbb{R}_{+}^{N \times M}$ as a matrix product of two non-negative matrices $\mathbf{W}$ and $\mathbf{H}$,
\begin{align}\label{eq:og_nmf}
    \mathbf{\tilde{V}} = \mathbf{W} \cdot \mathbf{H} \approx \mathbf{V}
\end{align}
where $\mathbf{W} \, \in  \mathbb{R}_{+}^{N \times K}$ and $\mathbf{H} \, \in  \mathbb{R}_{+}^{K \times M}$. $K$ is the rank of the decomposition.

Initially proposed in \cite{paatero1994positive, lee1999learning}, NMF soon saw innovative applications in audio \cite{smaragdis2003non, virtanen2007monaural}. There, NMF is most commonly applied to the STFT magnitude. Packing a magnitude spectrogram in a matrix and performing its NMF decomposition gives us a set of spectral template vectors as the columns of $\mathbf{W}$ and their corresponding activations as the rows of $\mathbf{H}$ \cite{smaragdis2003non}. As with any discrete transform, two crucially important parameters are the STFT window size, which defines the number of available frequency bins, and the hop size which determines the resulting number of spectra. Fixing these parameters determines the values of $N$ and $M$, which then determine the sizes of $\mathbf{W}$ and $\mathbf{H}$. The window size also determines the frequency resolution of the STFT.

To address varying tradeoffs between time and frequency resolution, the authors in \cite{rudoy2008adaptive,zhao2021optimizing} introduce adaptive approaches that can vary the window size dynamically to obtain adaptive resolution. Another way to trade off time and frequency resolution is to use the Constant-Q Transform (CQT) \cite{brown1991calculation,balazs2011theory}, or any other wavelet type transformation \cite{tzanetakis2001audio}, both of which result in representations that are not regularly sampled in either frequency or time. They cannot be represented in matrix form. Thus, we cannot perform NMF on these representations without rasterization or interpolation, which dilutes the initial benefits of these adaptive representations. \cite{zdunek2012approximation,hautecoeur2023least} talk about NMF with irregular samples, but they do so with a least-squares-based approach and not in the context of audio. We propose a data-driven framework using implicit neural representations \cite{sitzmann2020implicit} (INR) to model irregularly sampled time-frequency representations in a similar framework to NMF. In \Cref{sec:innmf}, we introduce INRs and show how they can factor irregular representations, and in \Cref{sec:compare}, we demonstrate that our proposed method performs equivalent to standard NMF.

\section{Implicit Neural NMF}
\label{sec:innmf}
We can think of time-frequency (T-F) representations as points in T-F space \cite{subramani2021point}, with the points being magnitudes indexed in terms of their underlying time-frequency coordinates. More specifically, we can consider $\mathcal{V}$ to be a set of $L$ time-frequency-magnitude tuples, $\mathcal{V} = \{(t_i,f_i, m_i)\}_{i = 1}^{L}$, where $m_i$ is the magnitude corresponding to time $t_i$ and frequency $f_i$. Given this data representation, we can express NMF in different terms as follows,
\begin{align}\label{eq:proposed_nmf}
    \tilde{m}_i = \sum_{k = 1}^{K} \mathcal{W}_k(f_i) \mathcal{H}_k(t_i) \approx m_i,
\end{align}
where $\{\mathcal{W}_k\}_{k = 1}^{K},\{\mathcal{H}_k\}_{k = 1}^{K}$ are the sets of spectral functions and their corresponding activation functions. The NMF approximation of a set of audio T-F points is thus the set of tuples $\tilde{\mathcal{V}} = \{(t_i,f_i, \tilde{m}_i)\}_{i = 1}^{L}$. \cite{smaragdis2013non} describe an approach to perform NMF on irregularly sampled data in terms of tuples. Our approach differs from theirs because here we will directly learn the templates and activations. The expression defined in \autoref{eq:proposed_nmf} has the advantage that it can be generalized for linear operations on arbitrarily sampled T-F representations.

If $\{f_i\}$ correspond to the regular DFT frequencies, and $\{t_i\}$ to the regular times, then $\{\mathcal{W}_k\}_{k = 1}^{K},\{\mathcal{H}_k\}_{k = 1}^{K}$ can be replaced by matrices $\mathbf{W},\mathbf{H}$, and \autoref{eq:proposed_nmf} collapses to the regular NMF equation for a magnitude spectrogram. However, since we are no longer constrained by regular sampling, the above formulation allows us to model the underlying basis templates and activations in a continuous space and sample these arbitrarily. Thus, we open up NMF to various representations beyond our regular STFT, namely the CQT (and other wavelet-based transformations), the sinusoidal models, and even the reassigned spectrogram \cite{flandrin2018time}. We are free to sample the T-F plane arbitrarily and query templates and activations at these locations. Our approach also side-steps the need to interpolate or resample all representations to a fixed one.

\begin{figure}[b]
    \centering
    \includegraphics[scale = 0.4]{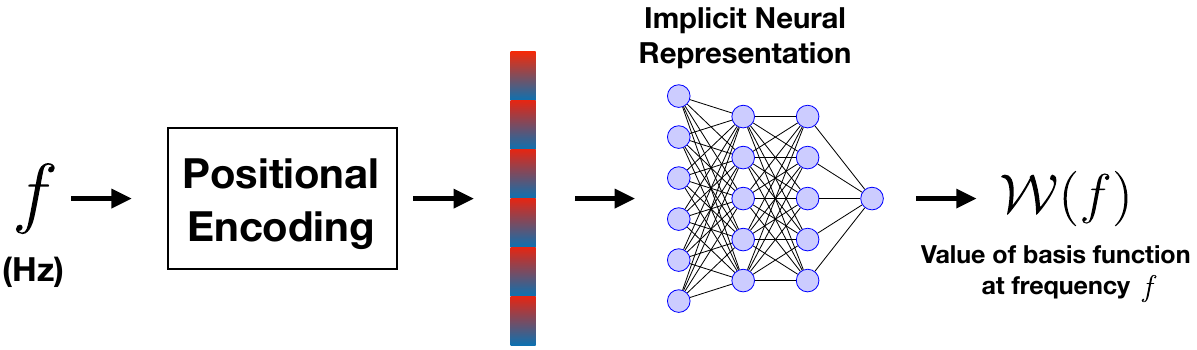}
    \caption{An Implicit Neural Representation describes $\mathcal{W}(.)$ as a function of frequency $f$ in Hz. Instead of directly giving $f$ as input, we obtain a Positional Encoding of $f$. The Implicit Neural Representation takes in this encoding and outputs the value of the basis function at that frequency $\mathcal{W}(f)$, thus giving us a function that we are free to sample at any arbitrary $f$. A similar network describes $\mathcal{H}(.)$}
    \label{fig:inr_demo}
\end{figure}

In order to operate in this new space, we make use of implicit neural representations and formulate our proposed model, implicit neural-NMF (iN-NMF), as an optimization problem. \cite{zdunek2012approximation} talks about an approach to decompose $\mathbf{W}$ into basis vectors, and then obtaining them via least-squares. \cite{hautecoeur2023least} also discuss approximating $\mathbf{W}$ with parametrized functions, and then obtaining the parameters with a least-squares approach. In contrast to their methods, we use machine learning models based on implicit representations to replace $W_k,H_k$ with learnable functions $\mathcal{W}_k,\mathcal{H}_k$.

\subsection{Implicit Neural Representations}
Implicit neural representations \cite{tancik2020fourier,sitzmann2020implicit} are a growing area of research in machine learning and signal processing. Synonymous with Neural Radiance Fields (NERF) \cite{mildenhall2021nerf}, these have been explored extensively in graphics and vision, and have currently begin to engender interest in the audio community \cite{10.1007/978-3-031-43421-1_39,NEURIPS2022_151f4dfc,NEURIPS2022_35d5ad98,lanzendorfer2023siamese, imamura2024neural}. The idea is to model signals by an underlying variable. In our case, we model each $\mathcal{W}_k$ and $\mathcal{H}_k$ as functions of frequency and time, respectively. These are called implicit representations because obtaining an `explicit' closed-form solution is challenging.

The idea with INRs is to give the underlying variable as an input to a neural network, and the output will be the function value. For example, instead of storing a value in the first column of $\mathbf{W}$ at index $i$ (which would correspond to the $i$-th frequency of our T-F representation), we can instead provide a real-valued number, corresponding to the Hz value of the $i$-th bin, to a neural network that will produce the same number. \autoref{fig:inr_demo} demonstrates this process.  This enables us to sample that column of $\mathbf{W}$ at any real-valued frequency without being constrained by a predetermined frequency spacing (or relying on simple interpolation schemes).

Instead of directly providing the underlying variable as a scalar input to a neural network, we found it best to use Fourier Encodings as described in \cite{tancik2020fourier,mildenhall2021nerf}. The Fourier encoding is a map $g:\mathbb{R} \rightarrow \mathbb{R}^{J}$ which projects the scalar input to a structured high-dimensional space (based on sinusoidal components in this case), which in practice works better than using a scalar input. The intermediate non-linearities that we use in our network are sinusoidal as described in \cite{sitzmann2020implicit}. To ensure non-negativity in the final outputs, the final non-linearity is a softplus. Thus, $\mathcal{W}_k$ and $\mathcal{H}_k$ are neural networks from $\mathbb{R} \rightarrow \mathbb{R}_{+}$ which can be queried anywhere in the input domain. For details of implementation and training, refer to our repository. For regularly-sampled representations, where the factors can be replaced with learnable matrices, it holds that $\mathcal{W}_k(f) = W_{f,k}$ and $\mathcal{H}_k(t) = H_{k,t}$.

\begin{algorithm}[b]
\label{alg:grad_innmf}
\caption{Gradient Descent for iN-NMF. \\ $\mathcal{W}_k,\mathcal{H}_k : \mathbb{R} \rightarrow \mathbb{R}_{+}$, $\mathcal{V} = \{(m_i,t_i,f_i)\}_{i = 1}^{L}$} \label{alg:in-NMF}
\begin{algorithmic}
\State $\theta$ \, \slash \slash \;Initially random parameters for $\{\mathcal{W}_k\}_{k = 1}^{K}$,$\{\mathcal{H}_k\}_{k = 1}^{K}$
\State $\eta$ \, \slash \slash \;Learning Rate
\State E \, \slash \slash \;Number of epochs
\For{$j=1$ to $E$}
    \For{$i=1$ to $L$}
    \State $\tilde{m}_i = \sum_{k = 1}^{K} \mathcal{W}_k(f_i) \mathcal{H}_k(t_i)$
    \State $\mathcal{L} = (m_i \log\frac{m_i}{\tilde{m}_i} - m_i + \tilde{m}_i)$
    \State $\theta = \theta - \eta \nabla \mathcal{L}$
    \EndFor
\EndFor
\end{algorithmic}
\end{algorithm}

\subsection{Formulating iN-NMF as an Optimization Problem}
Solutions to standard NMF can be found by minimizing a Kullback-Leibler(KL) like divergence between $\mathbf{V}$ and $\mathbf{\tilde{V}}$ \cite{lee2000algorithms}. In its original formulation in \cite{lee2000algorithms}, NMF uses a multiplicative update algorithm derived from a gradient descent approach with fine-tuned step sizes. For this paper, we will directly use a gradient-descent-based approach as described in \Cref{alg:in-NMF} for our proposed iN-NMF model. 

\begin{figure}[h]
    \centering
    \includegraphics[scale = 0.56]{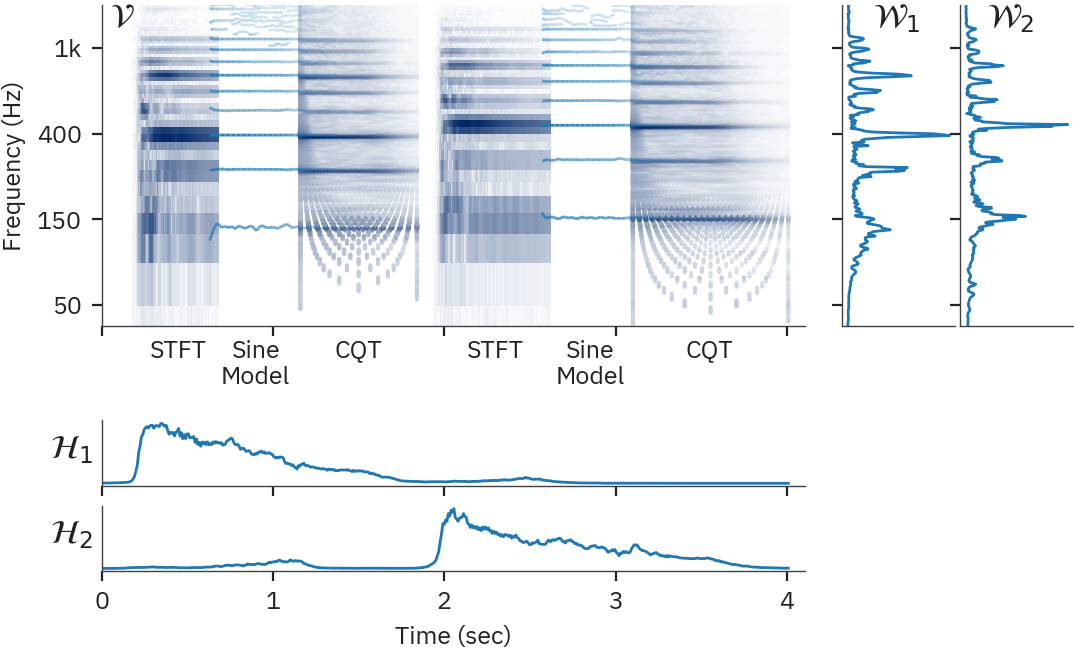}
    \caption{Decomposing two piano notes from a composite transform (top left image plot). For illustrative purposes, all the transforms are plotted on the same time-frequency axes, despite them all sampling that space differently. The learned factors from our algorithm are shown in the four line plots.}
    \label{fig:hybrid_demo}
\end{figure}

\subsection{Illustrative Example on a Hybrid Representation}
As an illustrative example that demonstrates the novel abilities of this approach, we show how we can find NMF-like bases and activations to describe a real-world signal that is represented using a sequence of varying T-F decompositions.\footnote{Of course in real life, one would not often encounter such a situation. This is just to demonstrate the flexibility that our approach affords us.} The input sound consists of two piano notes and is represented using an amalgam of three T-F representations for each note as shown in \autoref{fig:hybrid_demo}. The first transform is a short window STFT, the second is a sinusoidal model \cite{serra2013musical}, and the third is a constant-Q transform.  Had we used a single STFT representation and a rank-2 NMF, we would expect the two columns of $\mathbf{W}$ to converge to two spectra corresponding to the two notes; and consequently the rows of $\mathbf{H}$ would show us when these two notes are active.  However, given the hybrid representation we use here, this data cannot be appropriately packed as a time/frequency matrix, therefore it is impossible to decompose using standard techniques unless we resort to resampling and consequently annulling the benefits of each individual T-F transform.

We note that this representation is just a collection of magnitudes indexed by real-valued frequency and time values, and can be reduced to a list of \{time, frequency, magnitude\} tuples as described above.  Feeding these into our proposed model we obtain the functions shown in \autoref{fig:hybrid_demo}. The learned functions, as with NMF, do indeed reveal the spectrum of the two notes and when each note was active. They do so in a functional form that allows us to use these in alignment with any T-F decomposition.

\begin{figure}[h]
    \centering
    \includegraphics[scale = 0.11]{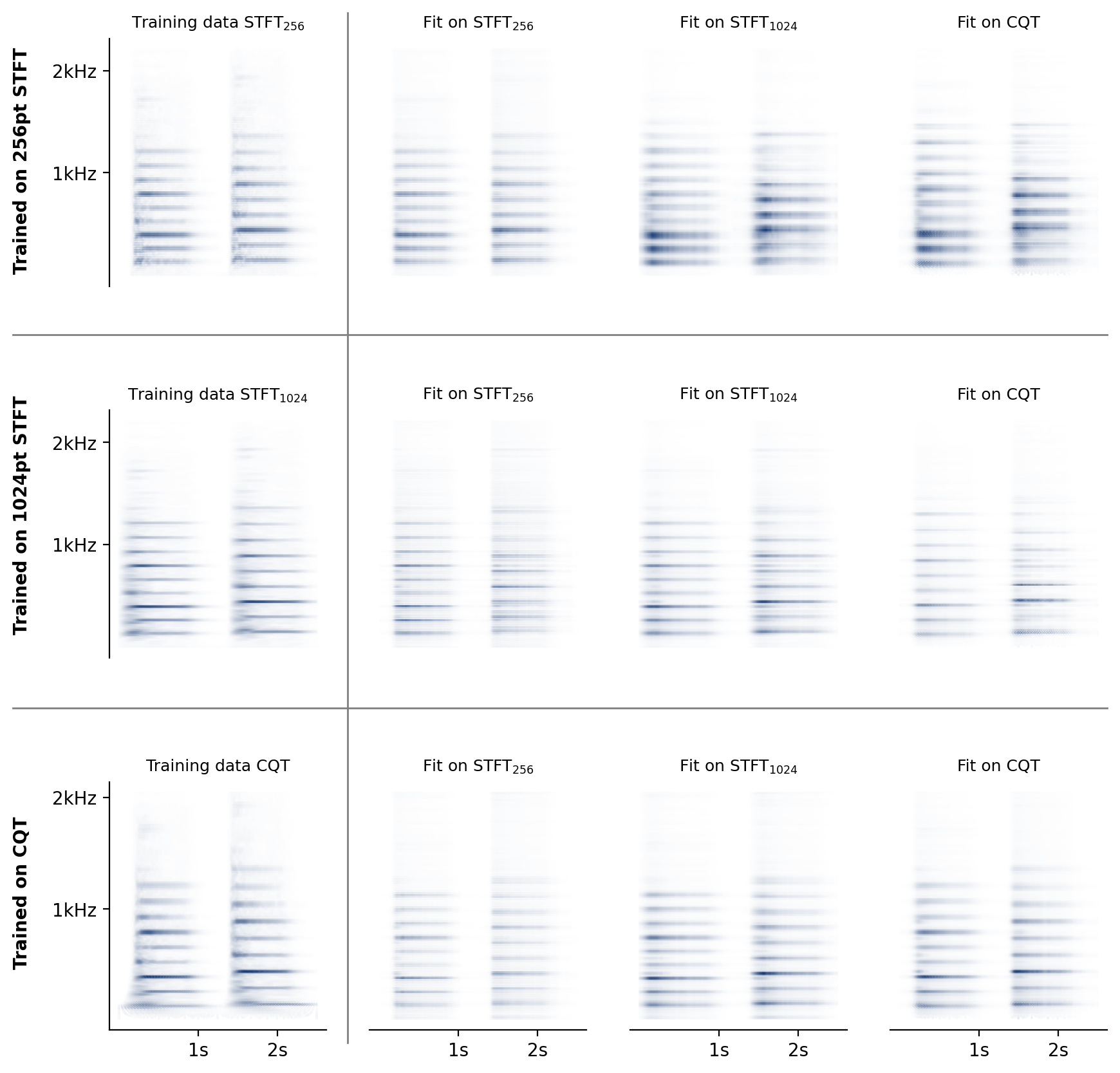}
    \caption{Testing generalization abilities.  Three models are trained on the different T-F representations shown in the leftmost column.  The spectral bases $\mathcal{W}$ of each, are then used to approximate the other two representations. Note that we plot all examples on the same frequency spacing to facilitate visual comparison. The 9 possible approximations are shown in the $3\times3$ grid on the right. Since our approach is not dependent on input representation, we see that the learned bases are able to explain all inputs well.}
    \label{fig:multiplot}
\end{figure}

To study the generalization capabilities of our learned factors, we construct the next experiment: we train separate models on different T-F representations individually (STFT with $N = 256$, STFT with $N = 1024$, and a CQT) and keep their respective $\mathcal{W}$'s. We then use each of these learned functions, but on the representation of the other two transforms, and estimate their respective activations. This is a common operation in NMF audio applications, where learned bases are used to approximate new inputs.  This setup will help us study how well our learned templates can generalize to an input represented using a different T-F transform. The results are shown in \autoref{fig:multiplot}.  The training data is shown in the three plots on the left. The entries in the nine right plots represent the approximations from each combination of training and test data. Since the factors we learn are continuous functions, we can sample them arbitrarily allowing us to use elements learned on one T-F representation to approximate a different T-F representation. Of course the quality of the learned function will be dependent on the training data.  For example, we note that the factors learned from the coarse resolution STFT (256 points), produce blurrier approximations along the frequency axis due to the lack of sufficient resolution during training. Regardless, all learned functions are able to approximate different T-F inputs as well as expected.

\section{Comparisons with NMF}
\label{sec:compare}

We will now present results on common NMF applications on speech data that show how the iN-NMF model compares.  We examine reconstructions of magnitude spectrograms given a learned dictionary, and NMF-based monophonic source separation.  In both cases we show that our proposed method is equivalent to matrix-based NMF.

\subsection{Magnitude Spectra Reconstruction}
We will begin with studying the reconstruction of magnitude spectrograms on the TIMIT dataset \cite{garofolo1993timit}. We will compare our model's performance to NMF with multiplicative updates \cite{lee2000algorithms} that minimize the KL-like divergence. We randomly sampled 500 different speakers and reconstructed their magnitude spectrograms after decomposing them with a factorization with $K = 20$. Since here we are working with spectrograms sampled regularly in time, we substitute $\{\mathcal{H}_k\}_{k = 1}^{20}$ with a learnable matrix $\mathbf{H}$ with no loss of generality. The learned spectral dictionaries $\{\mathcal{W}_k\}_{k = 1}^{20}$ will still be implicit neural representations.
We train our iN-NMF models using spectrograms with a DFT size of $N=2000$. We then use the same templates to also estimate activation matrices for spectrograms of different DFT sizes [1000, 1500, 2500]. This is similar to the approach discussed in \cite{smaragdis2017neural}, albeit this time we do so over varying spectrogram sizes. Doing so will show that our iN-NMF models can seamlessly generalize across different DFT sizes. Since matrix-based NMF is constrained to a single window size, we compare with multiple NMF models trained on each DFT size.
 \begin{table}[h]
 \centering
    \begin{tabular}{l|l|l}
    DFT size     & iN-NMF                      & NMF                         \\ \hline
    1000 & (51.89 $\pm$ 4.91)$\times 10^{-5}$ & (47.17 $\pm$ 4.28)$\times 10^{-5}$ \\ \hline
    1500 & (52.63 $\pm$ 4.65)$\times 10^{-5}$ & (48.82 $\pm$ 4.26)$\times 10^{-5}$ \\ \hline
    \rowcolor[gray]{.8}
    2000 & (52.34 $\pm$ 4.63)$\times 10^{-5}$ & (48.87 $\pm$ 4.42)$\times 10^{-5}$ \\ \hline
    2500 & (54.03 $\pm$ 4.52)$\times 10^{-5}$ & (49.25 $\pm$ 3.98)$\times 10^{-5}$ \\ 
    \end{tabular}
\caption{Final KL-divergences after convergence for both algorithms. We see that iN-NMF performs consistently on other window sizes despite only being trained on N=2000.}
\label{tab:recon_kld}
\end{table}

\autoref{tab:recon_kld} shows the KL-divergences after both iN-NMF and NMF have converged. We see that iN-NMF performs almost as well as NMF. We note that iN-NMF has only been trained on $N = 2000$, but it performs reasonably well for the other window sizes without re-learning the templates. Such a setting is useful when we have an input at a different resolution than the one iN-NMF was trained on. With NMF, we could not utilize the same basis vectors without some interpolation or resampling. However, for iN-NMF, we can sample the basis functions at the new resolution and re-learn the activations for that new resolution. This paves the way for building flexible models that are resolution and sample rate invariant.

\begin{algorithm}[t]
\caption{Separation with iN-NMF \\ $\mathcal{V}_{mix} = \{(m_i,t_i,f_i)\}_{i = 1}^{L}$} \label{alg:in-NMF_sep}
\begin{algorithmic}
\State Learn $\{(\mathcal{W}^{1})_{k}\}_{k = 1}^{K}$, $\{(\mathcal{W}^{2})_{k}\}_{k = 1}^{K}$ using Algorithm. \ref{alg:in-NMF}
\State $\theta$ \, \slash \slash \;Parameters($\{(\mathcal{H}^{1})_{k}\}_{k = 1}^{K},\{(\mathcal{H}^{2})_{k}\}_{k = 1}^{K})$
\State $\eta$ \, \slash \slash \;Learning Rate
\State E \, \slash \slash \;Number of epochs
\For{$j=1$ to $E$}
    \For{$i=1$ to $L$}
    \State $(\tilde{m}^{1})_{i} = \sum_{k = 1}^{K} (\mathcal{W}^{1})_{k}(f_i) (\mathcal{H}^{1})_{k}(t_i)$
    \State $(\tilde{m}^{2})_{i} = \sum_{k = 1}^{K} (\mathcal{W}^{2})_{k}(f_i) (\mathcal{H}^{2})_{k}(t_i)$
    \State $\tilde{m}_{i} = (\tilde{m}^{1})_{i} + (\tilde{m}^{2})_{i}$
    \State $\mathcal{L} = (m_i \log\frac{m_i}{\tilde{m}_i} - m_i + \tilde{m}_i)$
    \State $\theta = \theta - \eta \nabla \mathcal{L}$
    \EndFor
\EndFor
\end{algorithmic}
\end{algorithm}

\subsection{Monophonic Source Separation}
We also demonstrate the practical usability of our iN-NMF model in monophonic source separation. Again, we use the TIMIT dataset and construct 0 dB mixtures of two speakers. We use eight of the ten files of each individual speaker data to obtain spectral dictionaries, and two of the ten files to create mixtures.  We use both models to estimate the magnitude spectrograms $\tilde{\mathbf{V}}_i$, and 
the respective time-domain signals will be given by,
\begin{align}
    x_i(t) = \textrm{STFT}^{-1}(\frac{\tilde{\mathbf{V}}_i}{\sum_{j}\tilde{\mathbf{V}}_j} \odot \mathbf{V}_{mix}),
\end{align}
where $\mathbf{V}_{mix}$ is the STFT of the mixture. 

For NMF we first learn the basis from clean speech, and then learn activations for the mixture, as demonstrated in \cite{smaragdis2017neural}. Since we are comparing different window sizes, we have to re-run NMF whenever we change the window size. iN-NMF follows a similar procedure to the previous section. We train the templates on clean spectrograms with DFT size $N = 2000$, sample them for different window sizes, and learn latent representations for the two sources that explain the mixture, like in \cite{smaragdis2017neural}.  The procedure is explained in \Cref{alg:in-NMF_sep}. 

\begin{figure}[t]
    \centering
    \includegraphics[scale = 0.14]{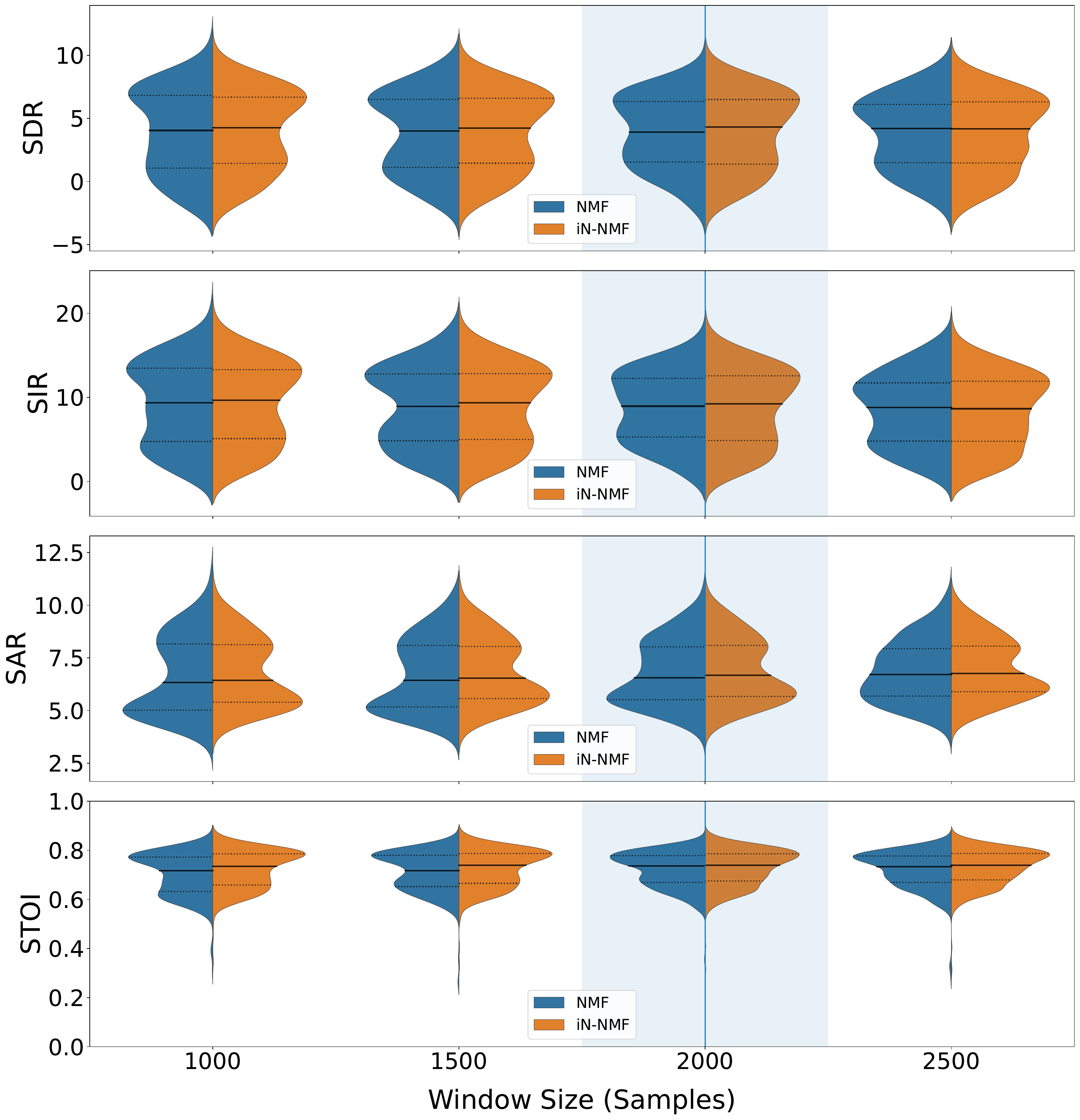}
    \caption{BSS\_Eval and STOI metrics comparing iN-NMF and NMF. We see that iN-NMF performs almost identically to NMF, even on inputs with different DFT sizes despite only being trained on a DFT size of $N = 2000$.}
    \label{fig:expt_separation}
\end{figure}

\autoref{fig:expt_separation} shows the BSS\_EVAL Metrics \cite{fevotte2005bss_eval} and STOI \cite{taal2011algorithm} for 500 random runs for NMF and iN-NMF across different window sizes. Even though iN-NMF has only been trained on $N=2000$, it performs consistently across all window sizes, almost as well as NMF, which must be run for every window size. Thus, we have a model that can be trained on a single window size and generalizes to a wide range of sizes.

\section{Conclusion}
We have introduced a new framework for linearly processing signals that are not regularly sampled. We demonstrate NMF on these classes of signals and show that our proposed model performs as well as conventional NMF but is more flexible to the input representation. We note here that our framework is not restricted to NMF; it can be generalized to any class of linear operations on signals.


\clearpage
\bibliographystyle{IEEEtran}
\bibliography{refs_nmf}

\begin{thebibliography}{10}
\providecommand{\url}[1]{#1}
\csname url@samestyle\endcsname
\providecommand{\newblock}{\relax}
\providecommand{\bibinfo}[2]{#2}
\providecommand{\BIBentrySTDinterwordspacing}{\spaceskip=0pt\relax}
\providecommand{\BIBentryALTinterwordstretchfactor}{4}
\providecommand{\BIBentryALTinterwordspacing}{\spaceskip=\fontdimen2\font plus
\BIBentryALTinterwordstretchfactor\fontdimen3\font minus \fontdimen4\font\relax}
\providecommand{\BIBforeignlanguage}[2]{{%
\expandafter\ifx\csname l@#1\endcsname\relax
\typeout{** WARNING: IEEEtran.bst: No hyphenation pattern has been}%
\typeout{** loaded for the language `#1'. Using the pattern for}%
\typeout{** the default language instead.}%
\else
\language=\csname l@#1\endcsname
\fi
#2}}
\providecommand{\BIBdecl}{\relax}
\BIBdecl

\bibitem{paatero1994positive}
P.~Paatero and U.~Tapper, ``Positive matrix factorization: A non-negative factor model with optimal utilization of error estimates of data values,'' \emph{Environmetrics}, vol.~5, no.~2, pp. 111--126, 1994.

\bibitem{lee1999learning}
D.~D. Lee and H.~S. Seung, ``Learning the parts of objects by non-negative matrix factorization,'' \emph{Nature}, vol. 401, no. 6755, pp. 788--791, 1999.

\bibitem{smaragdis2003non}
P.~Smaragdis and J.~C. Brown, ``Non-negative matrix factorization for polyphonic music transcription,'' in \emph{2003 IEEE Workshop on Applications of Signal Processing to Audio and Acoustics (IEEE Cat. No. 03TH8684)}.\hskip 1em plus 0.5em minus 0.4em\relax IEEE, 2003, pp. 177--180.

\bibitem{virtanen2007monaural}
T.~Virtanen, ``Monaural sound source separation by nonnegative matrix factorization with temporal continuity and sparseness criteria,'' \emph{IEEE transactions on audio, speech, and language processing}, vol.~15, no.~3, pp. 1066--1074, 2007.

\bibitem{rudoy2008adaptive}
D.~Rudoy, P.~Basu, T.~F. Quatieri, B.~Dunn, and P.~J. Wolfe, ``Adaptive short-time analysis-synthesis for speech enhancement,'' in \emph{2008 IEEE International Conference on Acoustics, Speech and Signal Processing}.\hskip 1em plus 0.5em minus 0.4em\relax IEEE, 2008, pp. 4905--4908.

\bibitem{zhao2021optimizing}
A.~Zhao, K.~Subramani, and P.~Smaragdis, ``Optimizing short-time {F}ourier transform parameters via gradient descent,'' in \emph{ICASSP 2021-2021 IEEE International Conference on Acoustics, Speech and Signal Processing (ICASSP)}.\hskip 1em plus 0.5em minus 0.4em\relax IEEE, 2021, pp. 736--740.

\bibitem{brown1991calculation}
J.~C. Brown, ``Calculation of a constant {Q} spectral transform,'' \emph{The Journal of the Acoustical Society of America}, vol.~89, no.~1, pp. 425--434, 1991.

\bibitem{balazs2011theory}
P.~Balazs, M.~D{\"o}rfler, F.~Jaillet, N.~Holighaus, and G.~Velasco, ``Theory, implementation and applications of nonstationary {G}abor frames,'' \emph{Journal of computational and applied mathematics}, vol. 236, no.~6, pp. 1481--1496, 2011.

\bibitem{tzanetakis2001audio}
G.~Tzanetakis, G.~Essl, and P.~Cook, ``Audio analysis using the discrete wavelet transform,'' in \emph{Proc. conf. in acoustics and music theory applications}, vol.~66.\hskip 1em plus 0.5em minus 0.4em\relax Citeseer, 2001.

\bibitem{zdunek2012approximation}
R.~Zdunek, ``Approximation of feature vectors in nonnegative matrix factorization with gaussian radial basis functions,'' in \emph{Neural Information Processing: 19th International Conference, ICONIP 2012, Doha, Qatar, November 12-15, 2012, Proceedings, Part I 19}.\hskip 1em plus 0.5em minus 0.4em\relax Springer, 2012, pp. 616--623.

\bibitem{hautecoeur2023least}
C.~Hautecoeur, L.~De~Lathauwer, N.~Gillis, and F.~Glineur, ``Least-squares methods for nonnegative matrix factorization over rational functions,'' \emph{IEEE Transactions on Signal Processing}, 2023.

\bibitem{sitzmann2020implicit}
V.~Sitzmann, J.~Martel, A.~Bergman, D.~Lindell, and G.~Wetzstein, ``Implicit neural representations with periodic activation functions,'' \emph{Advances in neural information processing systems}, vol.~33, pp. 7462--7473, 2020.

\bibitem{subramani2021point}
K.~Subramani and P.~Smaragdis, ``Point cloud audio processing,'' in \emph{2021 IEEE Workshop on Applications of Signal Processing to Audio and Acoustics (WASPAA)}.\hskip 1em plus 0.5em minus 0.4em\relax IEEE, 2021, pp. 31--35.

\bibitem{smaragdis2013non}
P.~Smaragdis and M.~Kim, ``Non-negative matrix factorization for irregularly-spaced transforms,'' in \emph{2013 IEEE Workshop on Applications of Signal Processing to Audio and Acoustics}.\hskip 1em plus 0.5em minus 0.4em\relax IEEE, 2013, pp. 1--4.

\bibitem{flandrin2018time}
P.~Flandrin, F.~Auger, and E.~Chassande-Mottin, ``Time-frequency reassignment: from principles to algorithms,'' in \emph{Applications in time-frequency signal processing}.\hskip 1em plus 0.5em minus 0.4em\relax CRC Press, 2018, pp. 179--204.

\bibitem{tancik2020fourier}
M.~Tancik, P.~Srinivasan, B.~Mildenhall, S.~Fridovich-Keil, N.~Raghavan, U.~Singhal, R.~Ramamoorthi, J.~Barron, and R.~Ng, ``Fourier features let networks learn high frequency functions in low dimensional domains,'' \emph{Advances in Neural Information Processing Systems}, vol.~33, pp. 7537--7547, 2020.

\bibitem{mildenhall2021nerf}
B.~Mildenhall, P.~P. Srinivasan, M.~Tancik, J.~T. Barron, R.~Ramamoorthi, and R.~Ng, ``Nerf: Representing scenes as neural radiance fields for view synthesis,'' \emph{Communications of the ACM}, vol.~65, no.~1, pp. 99--106, 2021.

\bibitem{10.1007/978-3-031-43421-1_39}
F.~Szatkowski, K.~J. Piczak, P.~Spurek, J.~Tabor, and T.~Trzci{\'{n}}ski, ``Hypernetworks build implicit neural representations of sounds,'' in \emph{Machine Learning and Knowledge Discovery in Databases: Research Track}, D.~Koutra, C.~Plant, M.~Gomez~Rodriguez, E.~Baralis, and F.~Bonchi, Eds.\hskip 1em plus 0.5em minus 0.4em\relax Cham: Springer Nature Switzerland, 2023, pp. 661--676.

\bibitem{NEURIPS2022_151f4dfc}
A.~Luo, Y.~Du, M.~Tarr, J.~Tenenbaum, A.~Torralba, and C.~Gan, ``Learning neural acoustic fields,'' in \emph{Advances in Neural Information Processing Systems}, S.~Koyejo, S.~Mohamed, A.~Agarwal, D.~Belgrave, K.~Cho, and A.~Oh, Eds., vol.~35.\hskip 1em plus 0.5em minus 0.4em\relax Curran Associates, Inc., 2022, pp. 3165--3177.

\bibitem{NEURIPS2022_35d5ad98}
K.~Su, M.~Chen, and E.~Shlizerman, ``Inras: Implicit neural representation for audio scenes,'' in \emph{Advances in Neural Information Processing Systems}, S.~Koyejo, S.~Mohamed, A.~Agarwal, D.~Belgrave, K.~Cho, and A.~Oh, Eds., vol.~35.\hskip 1em plus 0.5em minus 0.4em\relax Curran Associates, Inc., 2022, pp. 8144--8158.

\bibitem{lanzendorfer2023siamese}
L.~A. Lanzend{\"o}rfer and R.~Wattenhofer, ``Siamese siren: Audio compression with implicit neural representations,'' \emph{arXiv preprint arXiv:2306.12957}, 2023.

\bibitem{imamura2024neural}
K.~Imamura, T.~Nakamura, K.~Yatabe, H.~Saruwatari \emph{et~al.}, ``Neural analog filter for sampling-frequency-independent convolutional layer,'' \emph{APSIPA Transactions on Signal and Information Processing}, vol.~13, no.~1, 2024.

\bibitem{lee2000algorithms}
D.~Lee and H.~S. Seung, ``Algorithms for non-negative matrix factorization,'' \emph{Advances in neural information processing systems}, vol.~13, 2000.

\bibitem{serra2013musical}
X.~Serra, ``Musical sound modeling with sinusoids plus noise,'' in \emph{Musical signal processing}.\hskip 1em plus 0.5em minus 0.4em\relax Routledge, 2013, pp. 91--122.

\bibitem{garofolo1993timit}
J.~S. Garofolo, L.~F. Lamel, W.~M. Fisher, J.~G. Fiscus, D.~S. Pallett, N.~L. Dahlgren, and V.~Zue, ``{TIMIT} acoustic-phonetic continuous speech corpus,'' \emph{Philadelphia: Linguistic Data Consortium}, 1993.

\bibitem{smaragdis2017neural}
P.~Smaragdis and S.~Venkataramani, ``A neural network alternative to non-negative audio models,'' in \emph{2017 IEEE International Conference on Acoustics, Speech and Signal Processing (ICASSP)}.\hskip 1em plus 0.5em minus 0.4em\relax IEEE, 2017, pp. 86--90.

\bibitem{fevotte2005bss_eval}
C.~F{\'e}votte, R.~Gribonval, and E.~Vincent, ``{BSS}\_eval toolbox user guide--revision 2.0,'' \emph{[Technical Report] 2005, pp.19. ffinria-00564760}, 2005.

\bibitem{taal2011algorithm}
C.~H. Taal, R.~C. Hendriks, R.~Heusdens, and J.~Jensen, ``An algorithm for intelligibility prediction of time--frequency weighted noisy speech,'' \emph{IEEE Transactions on Audio, Speech, and Language Processing}, vol.~19, no.~7, pp. 2125--2136, 2011.

\end{thebibliography}







\end{document}